\documentclass[aps,pra,preprint,superscriptaddress]{revtex4-1}
\usepackage{graphicx}
\usepackage{amsfonts}
\usepackage{amsmath}
\usepackage{color}
\usepackage{bm}
\usepackage{amssymb}

\usepackage{multirow}
\usepackage{epstopdf}
\usepackage{natbib}

\newcommand{\beq}{\begin{equation}}
\newcommand{\eeq}{\end{equation}}

\def\pra#1#2#3{{Phys.~Rev.}~A~{\bf #1},\ #2\ (#3)}

\def\prl#1#2#3{{Phys.~Rev.~Lett.}~{\bf #1},\ #2\ (#3)}

\def\colvecnext#1{
        #1
        \global\advance\colveccount-1
        \ifnum\colveccount>0
                \\
                \expandafter\colvecnext
        \else
                \end{pmatrix}
        \fi
}

\begin{document}

\title{Atom-molecule collisions, spin relaxation, and sympathetic cooling in an ultracold spin-polarized Rb($^2\mathrm{S}$)-SrF$(^2\Sigma^+)$ mixture}
\author{Masato Morita}
\affiliation{Department of Physics, University of Nevada, Reno, NV, 89557, USA}
\author{Maciej B. Kosicki}
\affiliation{Institute of Physics, Astronomy and Informatics, Nicolaus Copernicus University, Toru{\'n}, Poland,  87-100}
\author{Piotr S. {\.Z}uchowski}
\affiliation{Institute of Physics, Astronomy and Informatics, Nicolaus Copernicus University, Toru{\'n}, Poland,  87-100}
\author{Timur V. Tscherbul}
\affiliation{Department of Physics, University of Nevada, Reno, NV, 89557, USA}

\pacs{}
\date{\today}

\begin{abstract}
We explore the suitability of ultracold collisions between spin-polarized SrF($^2\Sigma^+$) molecules and Rb($^2$S) atoms as elementary steps for the sympathetic cooling of SrF($^2\Sigma^+$) molecules in a magnetic trap. To this end, we carry out quantum mechanical scattering calculations on ultracold Rb~+~SrF collisions in a  magnetic field based on an accurate potential energy surface for the triplet electronic state of Rb-SrF developed {\it ab initio} using a spin-restricted coupler cluster  method with single, double and noniterative triple excitations [RCCSD(T)]. The Rb-SrF interaction has a global minimum with a well depth of 3444 cm$^{-1}$  in a bent geometry and a shallow local minimum  in the linear geometry.
Despite such a strong and anisotropic interaction, we find that converged close-coupling scattering calculations on Rb~+~SrF collisions in a magnetic field are still possible using rotational basis sets including up to 125 closed rotational channels in the total angular momentum representation.   Our calculations show that electronic spin relaxation in fully spin-polarized Rb-SrF collisions occurs much more slowly than elastic scattering over a wide range of magnetic fields (1-1000 G) and collision energies ($10^{-5}-10^{-3}$~K) suggesting good prospects of sympathetic cooling of laser-cooled SrF($^2\Sigma^+$) molecules with spin-polarized Rb($^2$S) atoms in a magnetic trap. 
We show that incoming $p$-wave scattering plays a significant role in ultracold collisions due to the large reduced mass of the Rb-SrF collision pair. The calculated magnetic field dependence of the inelastic cross sections at 1.4 $\mu$K displays a rich resonance structure including a low-field $p$-wave resonance, which suggests that external magnetic fields can be used to enhance the efficiency of sympathetic cooling in heavy atom-molecule mixtures.


\end{abstract}

\maketitle

\section{Introduction}


The production, trapping and manipulation of cold molecular gases is expected to make a major impact on chemical physics, quantum information processing,  quantum simulation, and fundamental tests of physics beyond the Standard Model \cite{njp09}. The ability to manipulate cold molecules with external electromagnetic fields is key to the wide range of their proposed applications \cite{MishaMolPhys13}.  External field-induced Stark and Zeeman energy shifts, while insignificant at thermal collision energies, become of major importance at ultralow temperatures, where  they can be used to activate or suppress reaction mechanisms \cite{JunARPC14,prl06,FaradayDiscuss09,njp09,RomanPCCP08}. A variety of ingenious mechanisms to control the reaction rates  have been demonstrated experimentally, including the use of Fermi statistics, long-range dipole-dipole interactions,  and external confinement to control the reaction KRb + KRb $\to$ K$_2$ + Rb$_2$ \cite{KRb10,KRbNaturePhys}. Recent theoretical work has explored the important roles of geometric-phase effects \cite{BrianBalaGeomPhase1}, quantum chaos \cite{QuantumChaos1,QuantumChaos2}, and electric-field-induced reactive scattering resonances  \cite{prl15} in ultracold chemical reactions.

Since its first experimental demonstration in 1998 \cite{Jonathan98}, magnetic trapping remains a key experimental technique for the production and trapping of cold molecular gases. Examples of molecular radicals trapped using this technique include CaH \cite{Jonathan98}, NH \cite{NHtrapping}, OH \cite{OHtrappingMeijerARPC,JunARPC14}, O$_2$ \cite{O2trappinEd1} and more recently, CH$_3$ \cite{CH3trapping} and SrOH \cite{Ivan16}.  
Latest experimental advances in laser cooling and molecular beam deceleration  have enabled magnetic and magneto-optical trapping of molecular ensembles at much lower temperatures than was previously possible. Laser-cooled samples of SrF($^2\Sigma^+$) and CaF($^2\Sigma^+$) molecules have been trapped at temperatures $\leq$400 microKelvin \cite{SrFnature14,SrFthesisJohnBarry,SrFprl16,CaFbelowDopplerLimit} and two-dimensional magneto-optical trapping of YO($^2\Sigma^+$) radicals has been accomplished \cite{JunYO}.

While extremely low compared to ambient or even cryogenic conditions  ($T=1-4$ K), milliKelvin temperatures are still too high for manipulating molecular interactions with external electromagnetic fields. The primary tool for such manipulation---the magnetic Feshbach resonance \cite{ChinReview}---requires collisions in a single partial-wave ($s$-wave) regime, which occur at temperatures well below 1 mK for most molecules.
Direct laser-cooling and molecular beam deceleration cannot reach such low temperatures  due to their intrinsic limitations (such as the Doppler limit \cite{SrFprl16}), so alternative cooling methods must be employed  to reach the ultracold regime  \cite{CaFbelowDopplerLimit}.

Sympathetic cooling is one such method, based on cooling atomic and molecular species by immersion in a gas of coolant atoms \cite{LaraRb-OHprl}. The method relies on elastic collisions to transfer momentum between the hot molecules and the coolant atoms and  has been successfully used  to cool fermionic K atoms \cite{Science01}, leading to the production of a quantum degenerate Fermi gas \cite{JosephDFG06}. Inelastic collisions are detrimental to the cooling process as they release the internal ({\it e.g.} Zeeman) energy of trapped molecules, leading to undesirable heating and trap loss \cite{njp09}. Spin-relaxation (or depolarization) collisions, which flip the electron spin of the molecule, represent a major inelastic channel for molecular radicals confined in permanent magnetic traps \cite{Roman04,jcp10,TutorialChapter2017}.  For optimal cooling, the ratio of the cross sections for elastic to spin relaxation collisions $\gamma$ should exceed 100 \cite{njp09,HutsonTarbutt}.

The search for atom-molecule combinations with favorable collisional properties for sympathetic cooling experiments has stimulated the development of molecular collision theory  in the presence of magnetic fields by Volpi and Bohn  \cite{VolpiBohn} and Krems and Dalgarno \cite{Roman04}. These pioneering  theoretical studies  focused on collisions with He atoms  and found that due to the low anisotropy of the molecule-He interaction, collision-induced spin relaxation of light $^2\Sigma$ and $^3\Sigma$ molecules with large rotational constants occurs much more slowly than elastic scattering, leading to the prediction that  NH($^3\Sigma^-$) radicals could be  magnetically trapped in cryogenic He buffer gas, which was later realized experimentally \cite{NHtrapping,NHisotopes}. 

Ultracold paramagnetic atoms (such as the alkali-metal atoms or atomic nitrogen) offer a viable alternative to cryogenic helium, which is unsuitable for sympathetic cooling of molecules below 100 mK due to  its vanishing vapor pressure.
However, theoretical studies found large inelastic relaxation rates in collisions of molecular radicals OH$(^2\Pi)$ and NH($^3\Sigma$) with ultracold Rb atoms, suggesting that the alkali-metal atoms would be much less suitable for sympathetic cooling of magnetically trapped molecules than the alkaline-earth atoms such as Mg \cite{Mg-NH} or atomic Nitrogen \cite{N-NH,PiotrN-NH} or H \cite {HutsonH-OH}, which  present significant experimental difficulties associated with either trapping or detection.

More recent quantum scattering studies have shown, however, that $^2\Sigma$ molecular radicals such as  CaH and SrOH have  low spin relaxation rates in collisions with ground-electronic-state Li($^2$S) atoms in their maximally spin-stretched Zeeman states, despite the triplet Li-CaH and Li-SrOH interactions being extremely strong and anisotropic \cite{pra11,pra17}. The suppression of spin relaxation is due to  the weak spin-rotation coupling among the molecular rotational levels involved in spin-flipping transitions \cite{RomanPRA03,pra11} and opens up the possibility of sympathetic cooling of $^2\Sigma^+$ molecules by ultracold Li atoms \cite{pra11,pra17}.

While atomic Li appears as a promising coolant for $^2\Sigma$ molecules,  quantum scattering calculations on Li-molecule collisions performed thus far neglected the chemical reaction between co-trapped molecules and Li atoms  ({\it e.g.} Li + CaH $\to$ LiH + Ca), which  are energetically allowed for many $^2\Sigma^+$ molecules of interest such as CaH \cite{pra11,JonathanLiCaH}, SrOH \cite{pra17}, and SrF \cite{PiotrJPCA17}. These reactions are often assumed to be forbidden for spin-polarized reactants by conservation of the total spin $S$ of the reaction complex \cite{prl06,jcp07,JacekPRA15}. However, model calculations show that $S$-changing intersystem crossing can occur at substantial rates even in fully spin-polarized atom-molecule collisions \cite{JanssenN-NH}, triggering rapid chemical reactions \cite{JonathanLiCaH}, which are  detrimental for sympathetic cooling.

 Fortunately, the chemical reactions of $^2\Sigma$ molecular radicals with heavier alkali-metal atoms, such as Rb~+~SrF $\to$ RbF~+~Sr are strongly endothermic \cite{PiotrJPCA17} and will therefore not occur at ultralow temperatures.  This consideration, together with recent numerical  simulations of cooling dynamics of CaF molecules in a microwave trap \cite{HutsonTarbutt} suggests that Rb could be a better choice than Li as the coolant atom.
  However, the collisional properties of $^2\Sigma$ molecular radicals with alkali-metal atoms heavier than Li remain unexplored due to the large densities of rovibrational states and strongly anisotropic atom-molecule interactions \cite{pra11,pra17,PiotrJPCA17,jpb16}, which have thus far precluded converged quantum scattering computations on these heavy systems. As a result, it remains unclear whether the ratio of elastic to inelastic collision rates in Rb-molecule collisions is large enough to allow for efficient sympathetic cooling.
 
In this work, we investigate ultracold collisions in a chemically non-reactive atom-molecule mixture  Rb-SrF using coupled-channel quantum scattering calculations based on an accurate {\it ab initio} PES of triplet symmetry. This system can be realized experimentally by co-trapping laser-cooled SrF($X^2\Sigma$) molecules \cite{SrFnature14,SrFprl16} with Rb atoms. We show that despite a high density of rovibrational states of SrF and a large number of partial waves involved in cold Rb~+~SrF collisions, it is possible to carry out converged coupled-channel (CC) calculations of elastic and spin relaxation cross sections using a recently developed total angular momentum representation for molecular collisions in magnetic fields \cite{jcp10}.
We find that the ratios of elastic to inelastic cross sections, while not as favorable as for Li collisions \cite{pra11,pra17}, are nevertheless fairly large ($\gamma>10$) over most of the collision energy and magnetic field ranges studied, with $\gamma>100$ reachable by tuning the external magnetic field and/or collision energy. We also find  a rich resonance structure in the spin relaxation cross sections as a function of applied magnetic field at ultralow collision energies ($1.4$ $\mu$K). The most of resonance structure arises due to the incoming $p$ partial-wave contributions, which are present even at very low collision energies due to the large reduced mass of the Rb-SrF collision complex, leading to a dramatic enhancement of the inelastic cross section. Our results suggest that the efficiency of sympathetic cooling in spin-polarized Rb-SrF($X^2\Sigma$) mixtures can be enhanced by tuning the spin relaxation cross sections away from resonance with an applied magnetic field.

This article is organized as follows. Section IIA presents our {\it ab initio} calculations of the triplet Rb-SrF potential energy surface (PES) and explores the main features of the PES. Sec. IIB outlines the methodology of Rb+SrF quantum scattering calculations in a magnetic field using the total angular momentum representation.  Section III presents the results for the elastic and inelastic cross sections as a function of collision energy and magnetic field, along with an analysis of the spin relaxation mechanisms. Section IV concludes with a summary of the main results and an outline of future research directions.




\section{Theory}

\subsection{Ab initio calculations of the triplet Rb-SrF PES}

As mentioned in the Introduction, the endothermicity of the chemical reaction  Rb + SrF $\to$ RbF + Sr  \cite{PiotrJPCA17} makes atomic Rb particularly attractive as a collision partner for sympathetic cooling of SrF. 
To pave the way for quantum dynamics calculations, we have carried out high-level {\it ab initio}  calculations  on the ${}^3A'$ electronic state of  Rb-SrF using the state-of-the-art  coupled cluster method with single, double and noniterative triple excitations [CCSD(T)] \cite{Hampel:92, Knowles:94} implemented in the MOLPRO package \cite{MOLPRO_brief:2015}. The augmented core-valence, correlation-consistent basis set (aug-cc-pCVQZ) was employed to describe the F atom. For the Rb and Sr atoms  the small-core relativistic energy-consistent pseudopotentials (ECP28MDF) were used together with tailored valence basis set \textit{spdfg}. All basis functions were uncontracted \cite{Lim:2005, Sr_ECP28MDF} and subsequently augmented by adding a single set of even-tempered functions. 
The interaction energy from the supermolecular calculations was counter-poise corrected to eliminate the basis-set superposition error (BSSE)\cite{Boys:70}. 
To describe the geometry of the Rb-SrF collision complex we use Jacobi coordinates $R$ and $\theta$, where $R$ is the distance between Rb and the center of mass of SrF and $\theta$ is the angle between the SrF axis  and the vector pointing from the center of mass  of SrF to Rb.  Throughout this paper, we assume that the SrF molecule is rigid and compute the interaction energy as a function of $R$ and $\theta$ at a fixed SrF bond length ($r=2.075$  \AA{}) corresponding to the  experimentally measured equilibrium geometry \cite{Huber}.

The potential is calculated on a dense two-dimensional grid of $\theta$  and $R$ extending from 2 to 10~\AA{} in steps of $\Delta\theta=5^\circ$ and $\Delta R=0.25$ \AA{}. For a given value of $R$ the potential is interpolated using the reproducing kernel Hilbert space (RKHS) method \cite{Ho:1996}. The RKHS parameters were set to extrapolate the interaction energy as 
$-C_6 R^{-6} - C_7 R^{-7} - C_8 R^{-8}$ beyond 10~\AA{} \cite{LaraRb-OHprl}.  We monitored the stability of the coupled-cluster calculations using the T1-diagnostics \cite{Lee:1989} with a result below 0.02 for all the $R$ gridpoints investigated.

The interpolated {\it ab initio} PES is expanded in Legendre polynomials as
\begin{equation}
V_\lambda(R)=\frac{1}{2}\left(2\lambda+1\right)\int_{-1}^{1} V(R,\theta) P_\lambda(\cos(\theta)) d\cos \theta.
\end{equation}
Following Ref. \cite{Janssen:2009}, the angular integration is performed using the quadratures which accurately reproduce the isotropic part of the potential $V_0(R)$. Due to a very strong potential anisotropy, we used a large number of expansion terms $V_\lambda(R)$ with $\lambda\le 25$. To enure the smoothness of the potential beyond 10 \AA{}, we used the van der Waals analytical expansion in inverse powers of $R$ applied to terms with   $\lambda=0\ldots4$. For higher-order Legendre components the potential was damped to zero at $R> 11$~\AA{}. A smooth connection between the {\em ab initio} PES at short range and the analytical  expansion at long-range was ensured by using the  switching function introduced by Janssen {\em et al.}~\cite{Janssen:2009} between 9 and 11 \AA{}.

\begin{table}
	\caption{ Convergence of the interaction energy with the basis set size at the global minimum ($R=4.1$ \AA{} $\theta=25^{\circ}$) and two saddle points at linear geometries:
	 Rb-F-Sr (with $R=4.25$ \AA{}) and Rb-Sr-F ($R=6.8$ \AA{}).  The energy unit is cm$^{-1}$.   }
	 \begin{tabular}{c|cccc|}
	                	&   aug-cc-pCVTZ  &   aug-cc-pCVQZ   & aug-cc-pCV5Z   & CBS \\ \hline
		 global minimum &  3329      &    3444    &   3485   &       3521          \\
		 Rb-F-Sr  saddle point &   3093    &    3243    &   3292   &    3337         \\
		 Rb-Sr-F saddle point &    170   &    172    &  173     &   175               \\
	 \end{tabular}
	\label{tabconv}
\end{table}

\begin{figure}[ht]
\begin{center}
\includegraphics[height=0.33\textheight,keepaspectratio]{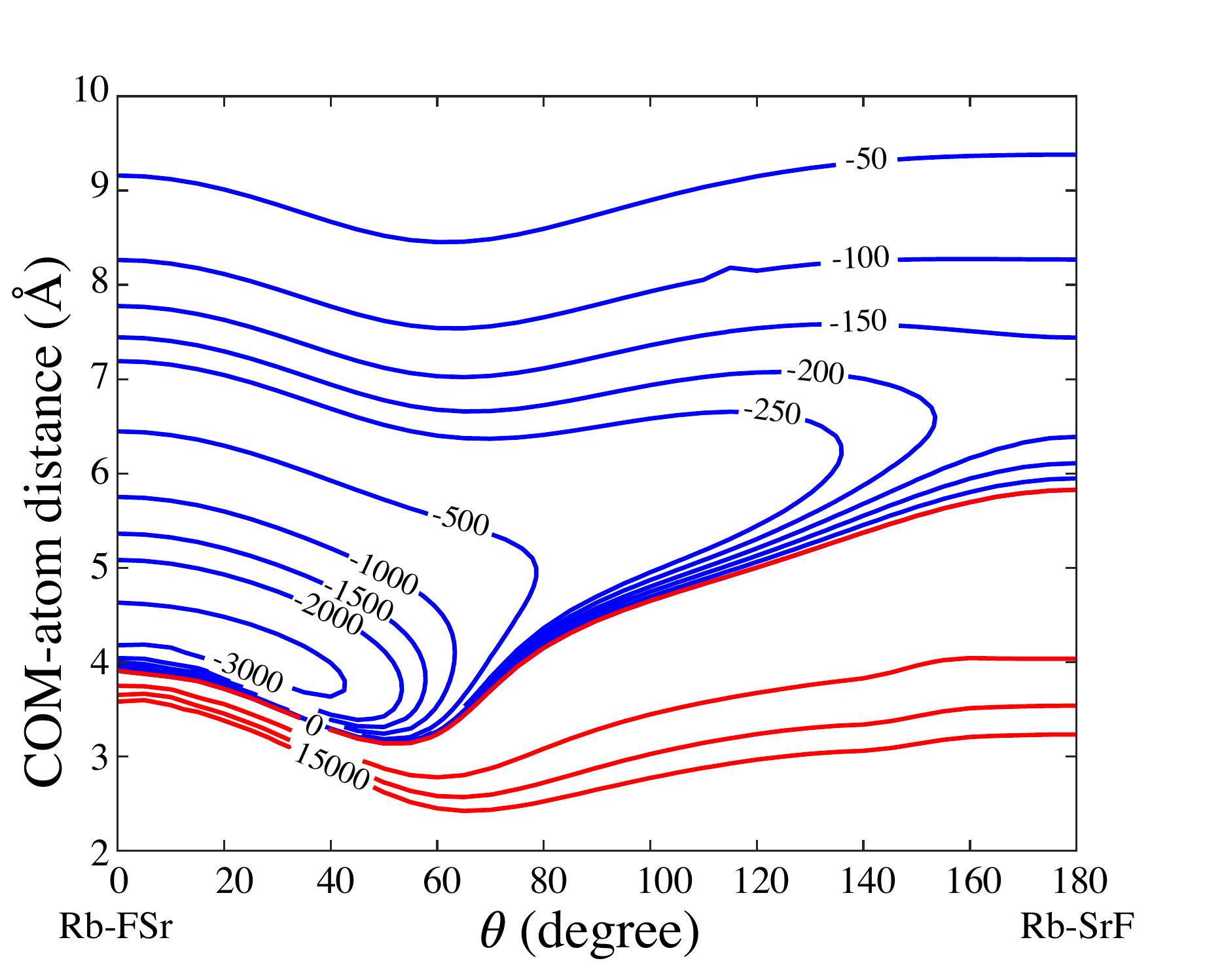}
\end{center}
\caption{\label{pes_fig} Contour plot of the {\it ab initio} potential energy surface for Rb-SrF in its triplet electronic state {(in units of cm$^{-1}$)}. The $\theta=0^{\circ}$ geometry corresponds to the collinear Rb--F-Sr arrangement. }
\label{figure1}
\end{figure}

A contour plot of the triplet Rb-SrF PES is shown in the Fig.~\ref{pes_fig}. The triplet PES at fixed $r$ corresponding to the equilibrium distance of SrF($X^2\Sigma$) has a global minimum in a bent configuration with $R = 4.1$\AA{}, $\theta=25^{\circ}$ with a well depth of $D_e=3444$ cm$^{-1}$. The potential is extremely anisotropic, leading one to expect strong coupling between the rotational states of SrF in the collision complex. There are two saddle points on the PES, both at linear geometries. For the Sr-F-Rb configuration the saddle point is located at $R=4.25$ \AA{}, while for the Rb-Sr-F configuration it is located at $R= 6.80$ \AA{}.
It is worthwhile to note that the  global minimum of the triplet PES is strongly attractive even at the restricted Hartree-Fock level of theory (about 2900 cm$^{-1}$ near the global minimum). This implies that the inaccuracy of our {\it ab initio} PES  should be smaller than that in typical dispersion-bound systems. To estimate the inaccuracy due to the incompleteness of the basis set,  we compare in Table~\ref{tabconv} the interaction energies near the stationary points of the PES obtained with series of basis sets of different quality, ranging from triple- to quintuple-zeta, as well as with the approximate complete basis set limit  (CBS).  Clearly, the depth of the potential near the global minimum and the  Rb-F-Sr saddle point changes very little with increasing basis set size. Moreover, we observe that  all the stationary points behave very similarly at the quintuple zeta level and in the CBS, so the shape of the PES is insensitive to the basis set. The global minimum obtained with the quadruple-zeta basis set, which was used in production calculations, is underestimated by 2.2\% compared to CBS limit. The corresponding figures for the  Rb-Sr-F and Rb-F-Sr saddle points are 1.7\% and 2.8\% respectively. Since the interaction energy of Rb-SrF is not dominated by the dispersion interaction, the contributions of higher excitations are marginal, we expect the CCSD(T) method to accurately reproduce the interaction energy.

\subsection{Quantum scattering calculations}
The quantum scattering problem for Rb~+~SrF in a magnetic field is solved by the numerical integration of close-coupling (CC) equations using the total angular momentum representation in the body-fixed (BF) coordinate frame \cite{jcp10,pra11}. We employ the rigid-rotor approximation by constraining the SrF bond length to the ground-state equilibrium value of $r=2.075$  \AA{}. The rigid-rotor approximation is justified by recent {\it ab initio} calculations \cite{PiotrJPCA17}, which have established that the Rb-SrF interaction  depends on $r$ only weakly.


The effective Hamiltonian for low-energy collisions between a $^2$S atom A (Rb) and a $^2\Sigma$ diatomic molecule B (SrF) in the presence of an external magnetic field may be written \cite{pra11,jcp10} 
\begin{equation}
\hat{\mathcal{H}} = - \frac{1}{2\mu} R^{-1} \frac{d^2}{dR^2} R
                     + \frac{(\hat{J}-\hat{N}-\hat{S}_{\mathrm{A}}-\hat{S}_{\mathrm{B}})^2}{2\mu R^2} \allowbreak
                     + \hat{\mathcal{H}}_{\mathrm{A}}
                     + \hat{\mathcal{H}}_{\mathrm{B}}
                     + \hat{\mathcal{H}}_{\mathrm{int}}
\label{eq:Heff}
\end{equation}
where A and B stand for Rb and SrF, $\mu$ is the reduced mass of the A-B collision complex 
$\mu=m_{\mathrm{A}}m_{\mathrm{B}}/(m_{\mathrm{A}}+m_{\mathrm{B}})$ with $m_{\mathrm{A}}=86.909180527$ and $m_{\mathrm{B}}=106.90401532$ a.m.u, $\hat{\mathcal{H}}_{\mathrm{A}}$ and $\hat{\mathcal{H}}_{\mathrm{B}}$ describe
non-interacting collision partners in an external magnetic field, and $\hat{\mathcal{H}}_{\mathrm{int}}$ is the atom-molecule interaction, which vanishes in the limit $R\to \infty$. The embedding of the BF $z$ axis is chosen to coincide with the vector $\boldsymbol{R}$, and the BF $y$ axis is chosen to be perpendicular to the plane defined by the collision complex.

In Eq. (\ref{eq:Heff}), $\hat{J}$ is the operator for the total angular momentum of the
collision complex, $\hat{N}$ is that for the rotational angular momentum of the diatomic molecule, and $\hat{S}_{\mathrm{A}}$ and $\hat{S}_{\mathrm{B}}$
are the operators for the electronic spin angular momenta of atom A and molecule B. The orbital angular momentum operator of the collision complex in the BF frame is given by $\hat{l} = (\hat{J}-\hat{N}-\hat{S}_{\mathrm{A}}-\hat{S}_{\mathrm{B}})$.
The Hamiltonian of atom A is given by $\hat{\mathcal{H}}_{\mathrm{A}}  = {{g}}_e\mu_\mathrm{B} \hat{S}_{{\rm
A},Z} B$, where ${g}_e$ is  the electron ${g}$-factor,
$\mu_\mathrm{B}$ is the Bohr magneton,  $\hat{S}_{{\rm A},Z}$ gives the projection
of $\hat{S}_{\rm A}$ onto the magnetic field axis, and $B$ is the magnitude of the external magnetic field. The Hamiltonian of the diatomic molecule B in its ground electronic state of $^2\Sigma$ symmetry  (such as SrF) is
\begin{equation}
\hat{\mathcal{H}}_{\mathrm{B}} = B_e
\hat{N}^2 + \gamma_\mathrm{SR} \hat{N} \cdot \hat{S}_{\rm B} + \textrm{g}_e\mu_\mathrm{B}
\hat{S}_{{\rm B},Z} B,
\end{equation}
where $B_e$ is the rotational constant, 0.2536135 cm$^{-1}$, and $\gamma_\mathrm{SR}=2.501\times 10^{-3}$ cm$^{-1}$ is the spin-rotation interaction constant. In this work, we neglect the weak hyperfine interactions due to the nuclear spins of $^{87}$Rb and $^{88}$Sr$^{19}$F for the sake of computational efficiency (adding these interactions would increase the number of channels by a factor of $(2I_\text{A}+1)\times (2I_\text{B}+1)=8$, increasing the computational cost over 100-fold). In the regime where the Zeeman splitting is small compared to the hyperfine interaction, scattering  calculations omitting the latter are known to underestimate the actual values of spin relaxation cross sections  \cite{N2}. The critical value of the magnetic field above which the hyperfine interactions become small compared to the Zeeman interaction (and hence can be neglected) is given by  $B_c = \Delta_{10}/\mu_\text{B}=77 $~G, where $\Delta_{10}=107.9$ MHz is the ground-state hyperfine splitting of $^{88}$Sr$^{19}$F ($I_\text{B}=1/2$) calculated using the molecular constants from Ref.~\cite{SrFthesisJohnBarry}.   Thus, while our results  at $B\geq 100$~G are likely to be only weakly affected by the hyperfine interaction, those at lower magnetic fields may be too small.

The atom-molecule interaction given by the $\hat{\mathcal{H}}_{\mathrm{int}}$ term in Eq. (\ref{eq:Heff}) 
includes both the electrostatic interaction potential $\hat{V}$ and the magnetic dipole-dipole 
interaction $\hat{{V}}_\mathrm{dd}$ between the magnetic moments of
the atom and the molecule. The interaction potential $\hat{V}$ may be written
\begin{equation}
 \hat{V}(R,\theta) = \sum^{S_{\mathrm{A}}+S_{\mathrm{B}}}_{S=|S_{\mathrm{A}}-S_{\mathrm{B}}|}
  \sum^{S}_{\Sigma=-S} |S \Sigma \rangle \hat{V}^S(R,\theta) \langle S \Sigma|\,,%
 \label{eq:V}
\end{equation}
where total electronic spin $S$ is defined as $\hat{S} =\hat{S}_{\mathrm{A}}+\hat{S}_{\mathrm{B}}$. 
In this work, we are interested in collisions between rotationally ground-state SrF molecules ($N=0$) with Rb atoms initially in their maximally stretched, magnetically trappable Zeeman states, {\it i.e.} $M_{S_\mathrm{A}}=M_{S_\mathrm{B}}=1/2$, where $M_{S_\mathrm{A}}$ and $M_{S_\mathrm{B}}$ are the projections of $\hat{S}_{\mathrm{A}}$ and $\hat{S}_{\mathrm{B}}$ onto the space-fixed $Z$-axis.
Following our previous work on Li-CaH and Li-SrOH \cite{pra11,pra17} we assume  that the non-adiabatic coupling between the triplet ($S = 1$) and the singlet ($S = 0$) Rb-SrF PESs can be neglected, and that the PESs are identical, {\it i.e.}  $\hat{V}^{S=0}(R,\theta)=\hat{V}^{S=1}(R,\theta)$.
The dipolar interaction between the magnetic moments of the atom and molecule
may be written \cite{jcp12a}
\begin{equation}
 \hat{V}_\mathrm{dd}=
 -\textrm{g}_{e}^{2}\mu_{0}^{2}\sqrt{\frac{24\pi}{5}} \frac{\alpha ^2}{R^3}  \sum^{}_{q} (-)^qY_{2,-q}^*(\hat{\boldsymbol{R}}) [\hat{S}_{\rm A}\otimes \hat{S}_{\rm B}]_{q}^{(2)},
 \label{eq:Hdip}
\end{equation}
where   $\mu_0$ is the magnetic permeability of free space, $\alpha$ is the fine-structure constant and $ [\hat{S}_{\rm A}\otimes \hat{S}_{\rm B}]_{q}^{(2)}$ is the spherical tensor product of $\hat{S}_{\rm A}$ and $\hat{S}_{\rm B}$.

Following previous theoretical work \cite{pra11,jcp10,jcp12a},  the total wave function of Rb-SrF collision complex is expanded in a set of basis functions 
\begin{equation}\label{eq:totjbasis}
\left|JM\Omega\rangle|NK_N\rangle|S_{\mathrm{A}}\Sigma_{\mathrm{A}}\rangle|S_{\mathrm{B}}\Sigma_{\mathrm{B}}\right\rangle.
\end{equation}
Here, $\Omega$, $K_N$, $\Sigma_{\mathrm{A}}$ and $\Sigma_{\mathrm{B}}$ are the projections of $J$, $N$, $S_{\mathrm{A}}$ and $S_{\mathrm{B}}$ onto the BF quantization axis $z$, and  $\Omega=K_N+\Sigma_{\mathrm{A}}+\Sigma_{\mathrm{B}}$. The projection of $J$ onto the space-fixed quantization axis $M$ is rigorously conserved for collisions in a static magnetic field \cite{Roman04,jcp10}, and we solve the CC equations separately for each value of $M$. In Eq.~(\ref{eq:totjbasis})  $|JM\Omega\rangle=\sqrt{(2J+1)/8\pi^2}D^{J*}_{M\Omega}(\bar{\alpha},\bar{\beta},\bar{\gamma})$ is an eigenfunction of the symmetric top, and the Wigner $\mathit{D}$-functions $D^{J*}_{M\Omega}(\bar{\alpha},\bar{\beta},\bar{\gamma})$ depend on the Euler angles $\bar{\alpha}$, $\bar{\beta}$ and $\bar{\gamma}$, which specify the position of the BF axes $x$, $y$ and $z$ in the SF frame. The rotational degrees of freedom of SrF in the BF frame are described by the functions $|NK_N\rangle$, which can be expressed using the spherical harmonics as $\sqrt{2\pi} Y_{NK_N}(\theta,0)$. The matrix elements of the effective Hamiltonian in the total angular momentum representation (\ref{eq:totjbasis}) are evaluated as described elsewhere \cite{jcp10}. The matrix elements of the magnetic dipole-dipole interaction $\hat{{V}}_\mathrm{dd}$ are given by
\begin{multline}
\langle J'M\Omega' | \langle N'K_N'| \langle S_{A}\Sigma_{A}'| \langle S_{B}\Sigma_{B}'|\hat{\mathrm{V}}_\mathrm{dd}|S_{B}\Sigma_{B}\rangle |S_{A}\Sigma_{A}\rangle |NK_N\rangle |JM\Omega \rangle 
          =  \delta _{J'^{}J}\delta _{\Omega' \Omega} \delta _{N' N} \delta _{N_K' N_K}  \\
           \times  \left(-\frac{\sqrt{30} \textrm{g}_{e}^{2}\mu_{0}^{2}\alpha^2}{R^3}\right)   
          (-1)^{S_A+S_{B}-\Sigma^{}_{A}-\Sigma^{}_{B}}  
          \sqrt{(2S_A+1)S_A(S_A+1)} \sqrt{(2S_{B}+1)S_{B}(S_{B}+1)}          \\
           \times \sum_{q_A, q_{B} } 
          \begin{pmatrix} 1 & 1 & 2\\ q_A & q_{B} & 0 \end{pmatrix}    
          \begin{pmatrix} S_A & 1 & S_A\\ -\Sigma^{'}_{A} & q_A & \Sigma^{}_{A} \end{pmatrix}
          \begin{pmatrix} S_{B} & 1 & S_{B}\\ -\Sigma^{'}_{B} & q_{B} & \Sigma^{}_{B} \end{pmatrix}. 
\label{aeq:Vdd}
\end{multline}

The size of the basis set is determined by the truncation parameters of $J_\mathrm{max}$ and $N_\mathrm{max}$ which give the maximum quantum numbers of the  total angular momentum $J$ of the collision complex Rb-SrF and the rotational angular momentum $N$ of SrF in the basis set. We explore the convergence of the cross sections with respect to these parameters in the Appendix, and all calculations in the text are performed with the converged values of $J_\mathrm{max}=3$ and $N_\mathrm{max}=125$. The numerical procedures used in this work are essentially the same as those employed in our previous studies of Li~+~CaH and Li~+~SrOH collisions \cite{pra11} and explained in detail elsewhere \cite{jcp10,jcp12a}. The CC equations are solved numerically using the log-derivative propagator method  \cite{prop_1, prop_2} on an equidistant radial grid from $R_\text{min}=5.2$ Bohr to $R_\text{mid}$ with $R_\text{mid}=15.0$ Bohr for $B \ge 10$~G and $R_\text{mid}=25.0$ Bohr for $B < 10$~G using a step size of $0.002$ Bohr. Airy propagation is employed for $R_\text{mid} \le R \le R_\text{max}$ with $R_\text{max}=300.0$ Bohr for $B \ge 10$~G and $R_\text{max}=750$~Bohr for $B < 10$~G.

\section{Results}

\subsection{Elastic and inelastic cross sections}
\begin{figure}[ht]
\begin{center}
\includegraphics[height=0.33\textheight,keepaspectratio]{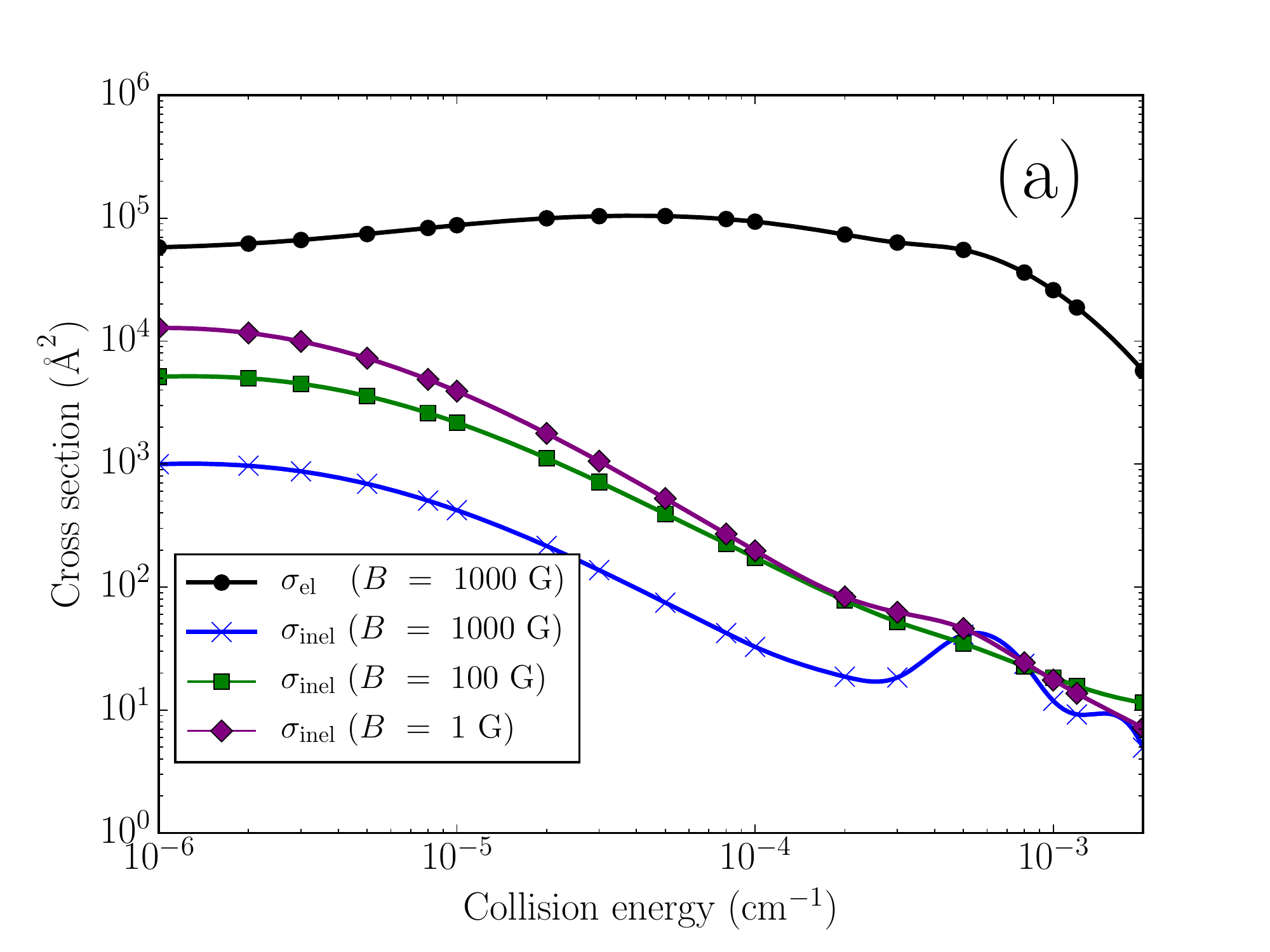}
\includegraphics[height=0.33\textheight,keepaspectratio]{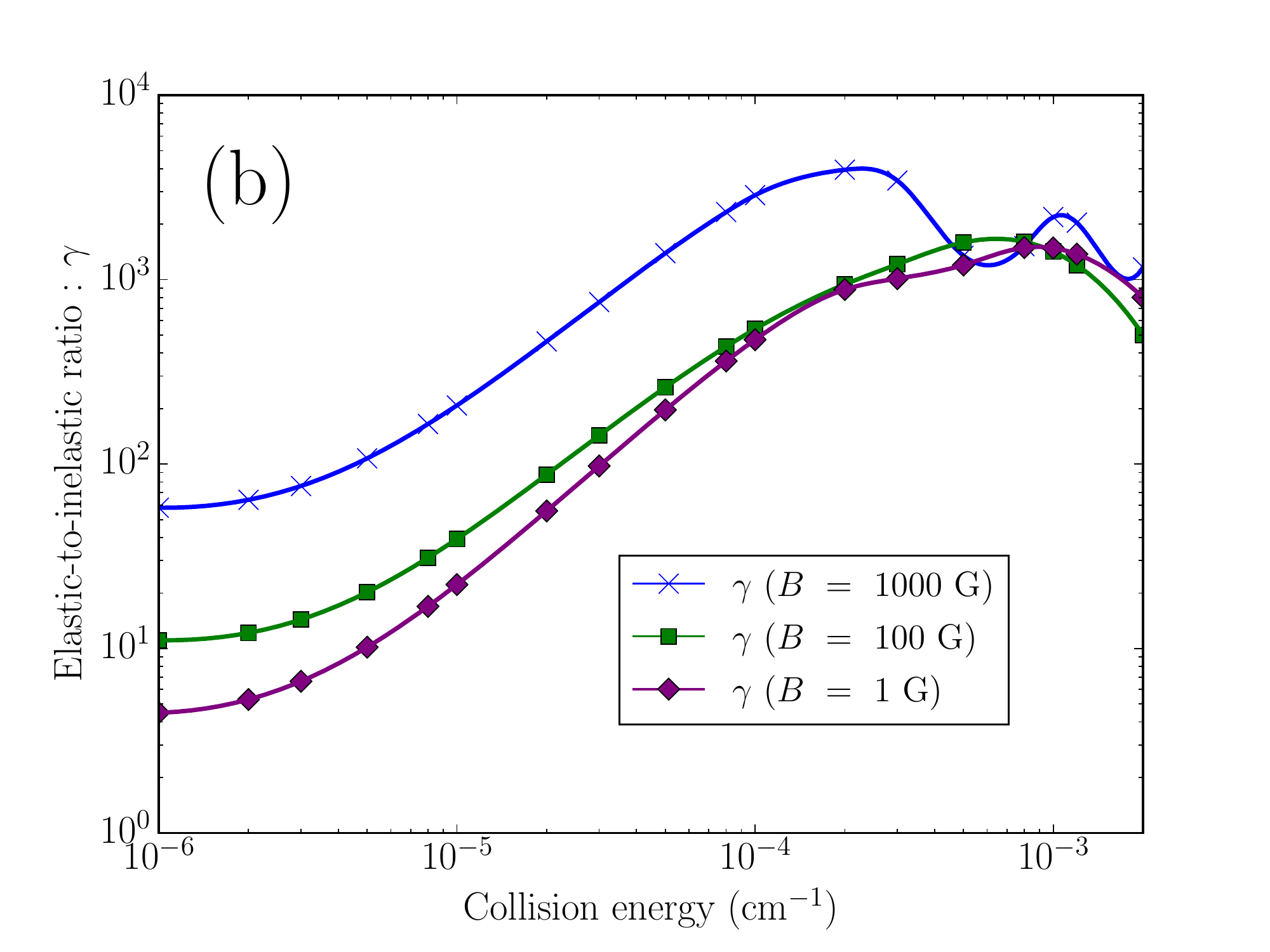}
\end{center}
\caption{ (a) Collision energy dependence of the elastic cross section (circles) and inelastic cross section for the external magnetic field of 1 G (diamonds), 100 G (squares) and 1000 G (crosses). The elastic cross section displays a very weak field dependence. (b) The ratios of elastic and inelastic cross sections as functions of collision energy for the same values of the magnetic field as in (a). }
\label{figure4}
\end{figure}

Figure \ref{figure4}(a) shows  the elastic and inelastic cross sections for spin-polarized Rb~+~SrF collisions plotted as functions of collision energy for the external magnetic fields of 1, 100, and 1000 G. The internal state of SrF($X^{2}\Sigma^+$) before the collision is $|N=0,M_N=0,M_{S_B}=1/2\rangle$  and that of  Rb($^2$S) is $|M_{S_A}=1/2\rangle$ in the space-fixed coordinate frame representation. At very low collision energies of interest here (which are much smaller than the rotational energy splitting between the ground $N=0$ and the first excited, $N=1$ rotational states of SrF), the only inelastic process that can occur is electronic spin relaxation within the ground rotational state, i.e. $|N=0,M_N=0,M_{S_B}=1/2\rangle \to |N'=0,M_N'=0,M_{S_B}'=-1/2\rangle$.
 The field dependence of the elastic cross section is very weak, and thus only  the $B=1000$~G result  is shown in Fig.~\ref{figure4} (a). We observe that the inelastic cross section decreases with increasing the magnetic field from 1~G to 1000~G; the effect is particularly strong  in the ultracold  $s$-wave regime.

A key figure of merit for sympathetic cooling is the ratio of elastic to inelastic cross sections $\gamma=\sigma_\mathrm{el}/\sigma_\mathrm{inel}$; $\gamma>100$ is generally required for optimal cooling of magnetically trapped molecules. Figure~\ref{figure4} (b) shows that the calculated values of $\gamma$ for Rb~+~SrF collisions exceed 100 at collision energies above $E_C\sim 5\times 10^{-5}$ cm$^{-1}$, suggesting good prospects for sympathetic cooling of cold SrF($^2\Sigma^+$) molecules with magnetically co-trapped Rb atoms.
At ultralow collision energies ($E_C<10^{-5}$~cm$^{-1}$)  the ratio of elastic to inelastic collision rates drops below 100 and becomes very sensitive to the applied magnetic field.  Still, we observe that the inelastic cross sections are relatively small at $B=1000$~G compared with  their values at smaller magnetic fields. Thus, as noted previously for He~+~O$_2$ \cite{He-O2} and Li~+~SrOH \cite{pra17}, it may be possible to enhance the efficiency of sympathetic cooling by tuning the inelastic cross sections with an applied magnetic field. 

\begin{figure}[ht]
\begin{center}
\includegraphics[height=0.35\textheight,keepaspectratio]{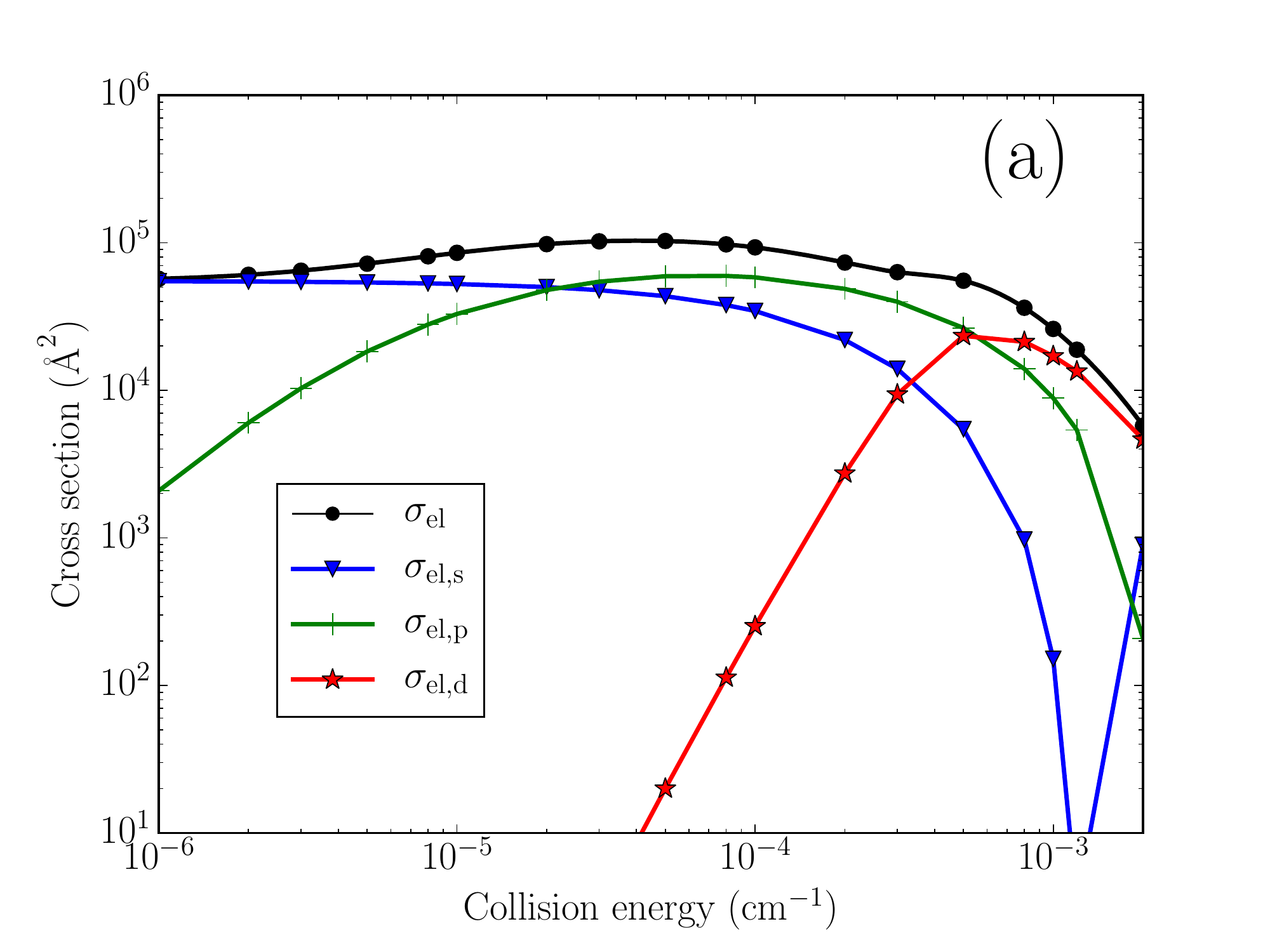}
\includegraphics[height=0.35\textheight,keepaspectratio]{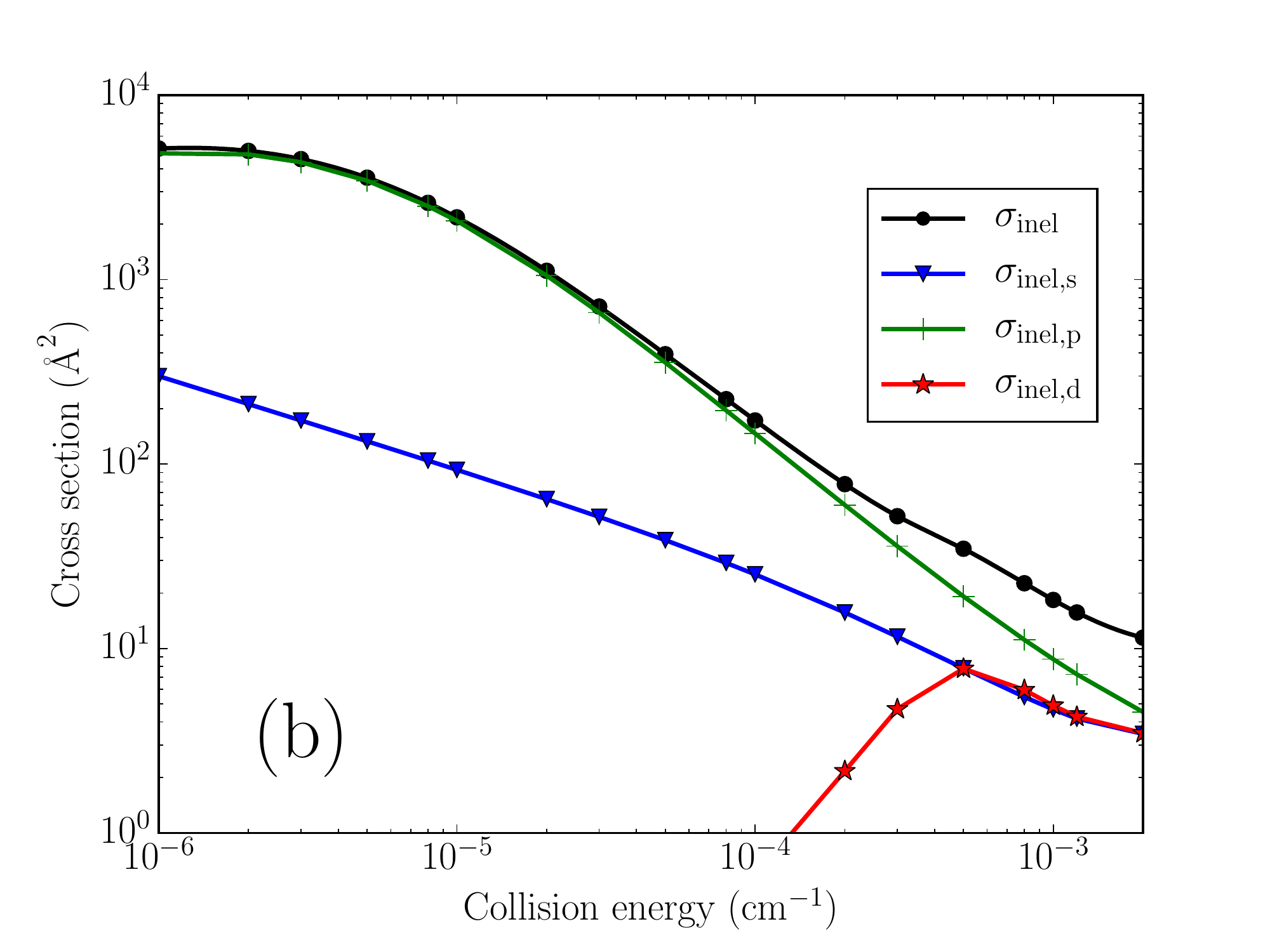}
\end{center}
\caption{Incoming partial wave decomposition of the elastic (a) and inelastic (b) cross sections as a function of collision energy calculated for the magnetic field of 100 G. The total elastic and inelastic cross sections are shown as solid lines with circles.  }
\label{figure5} 
\end{figure}

Figures\ \ref{figure5}(a)-(b) show incoming partial wave contributions to the elastic and inelastic cross sections at $B=100$~G.  Based on the {\it ab initio} value of the long-range dispersion coefficient $C_6=3495$ a.u. \cite{PiotrJPCA17}, the calculated heights of the $p$ and $d$-wave centrifugal barriers are $5.53\times 10^{-5}$ and and $2.87\times 10^{-4}$ cm$^{-1}$. Consistent with these estimates, we observe in  Fig.~\ref{figure5}(a) a decline of $l\ge 1$ incoming partial wave contributions to the elastic cross section as the collision energy is tuned below the corresponding barrier heights.

 Remarkably, the  $p$-wave contribution to the {\it inelastic} cross section dominates through the entire collision energy range spanning 3 orders of magnitude ($E_C= 10^{-6} - 2\times 10^{-3} $  cm$^{-1}$).  
 This suggests the presence of a near-threshold scattering resonance, as discussed in more detail below. In contrast, the partial wave spectrum of the inelastic cross sections calculated previously for  Li~+~CaH and Li~+~SrOH \cite{pra17} is dominated by the incoming $s$-wave contributions below $E_C=10^{-3}$~cm$^{-1}$ and by all partial waves at higher collision energies.



\subsection{Magnetic field dependence and spin-relaxation mechanisms}

\begin{figure}[ht]
\begin{center}
\includegraphics[height=0.28\textheight,keepaspectratio]{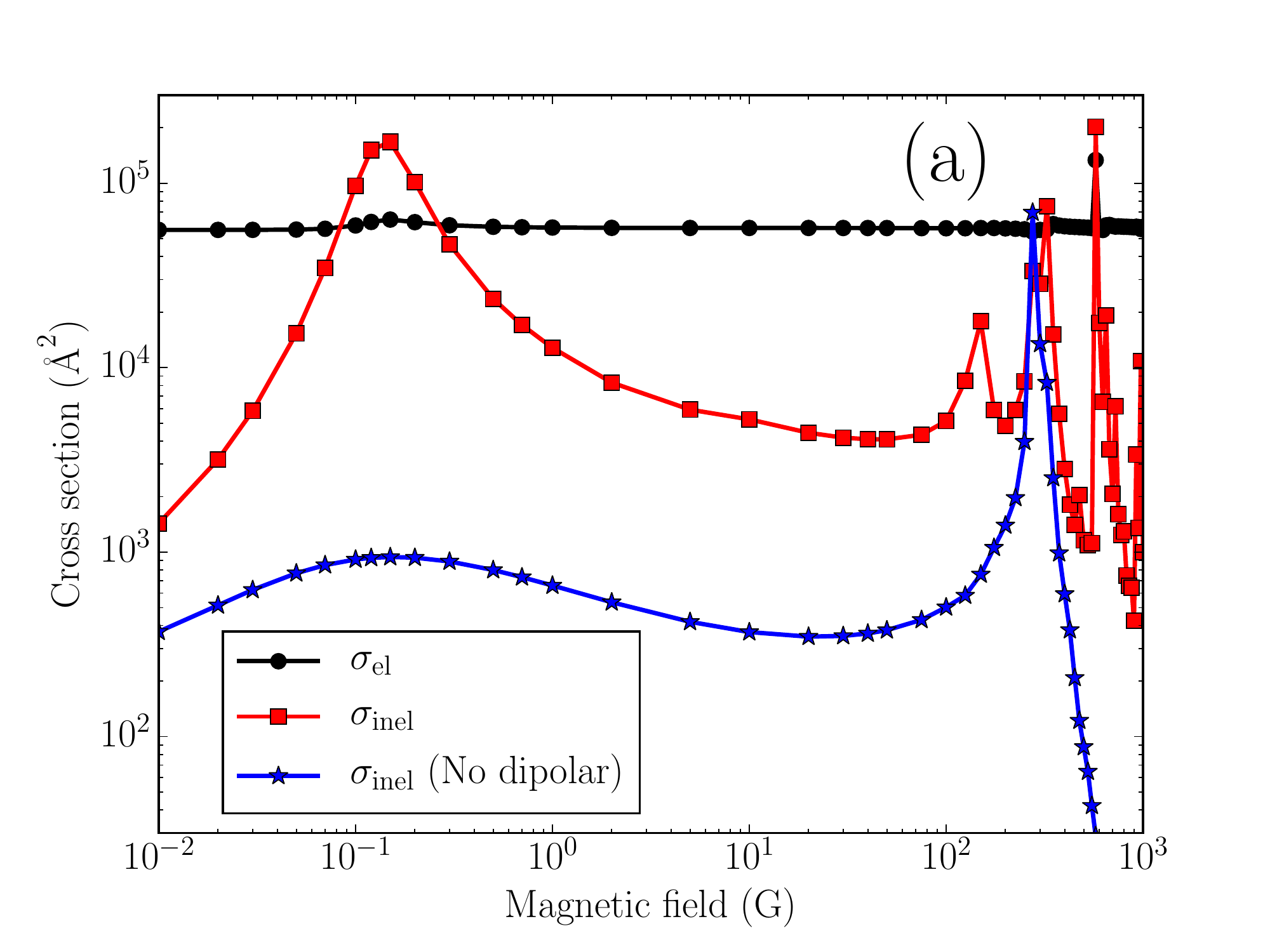}
\includegraphics[height=0.28\textheight,keepaspectratio]{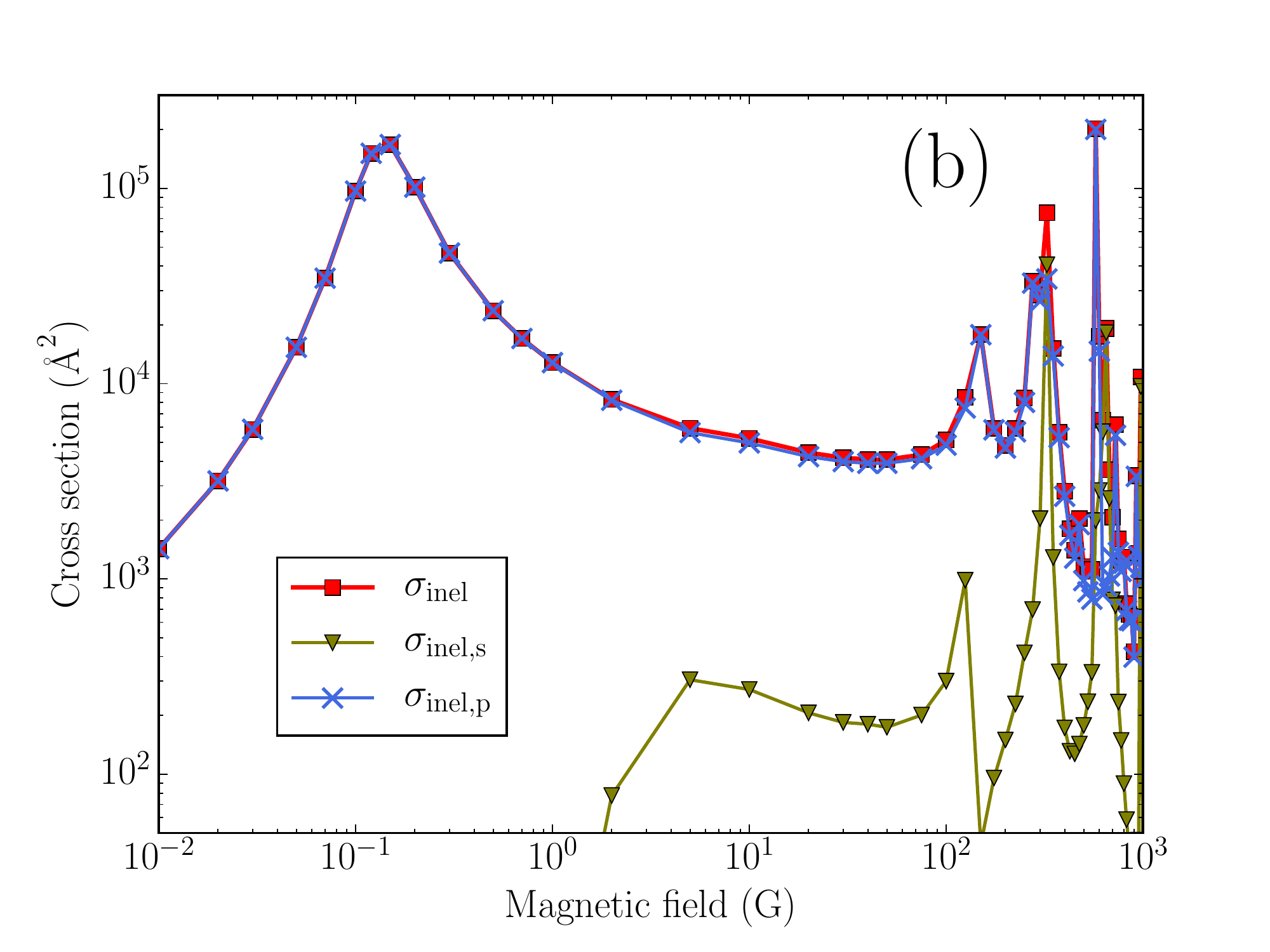}
\includegraphics[height=0.28\textheight,keepaspectratio]{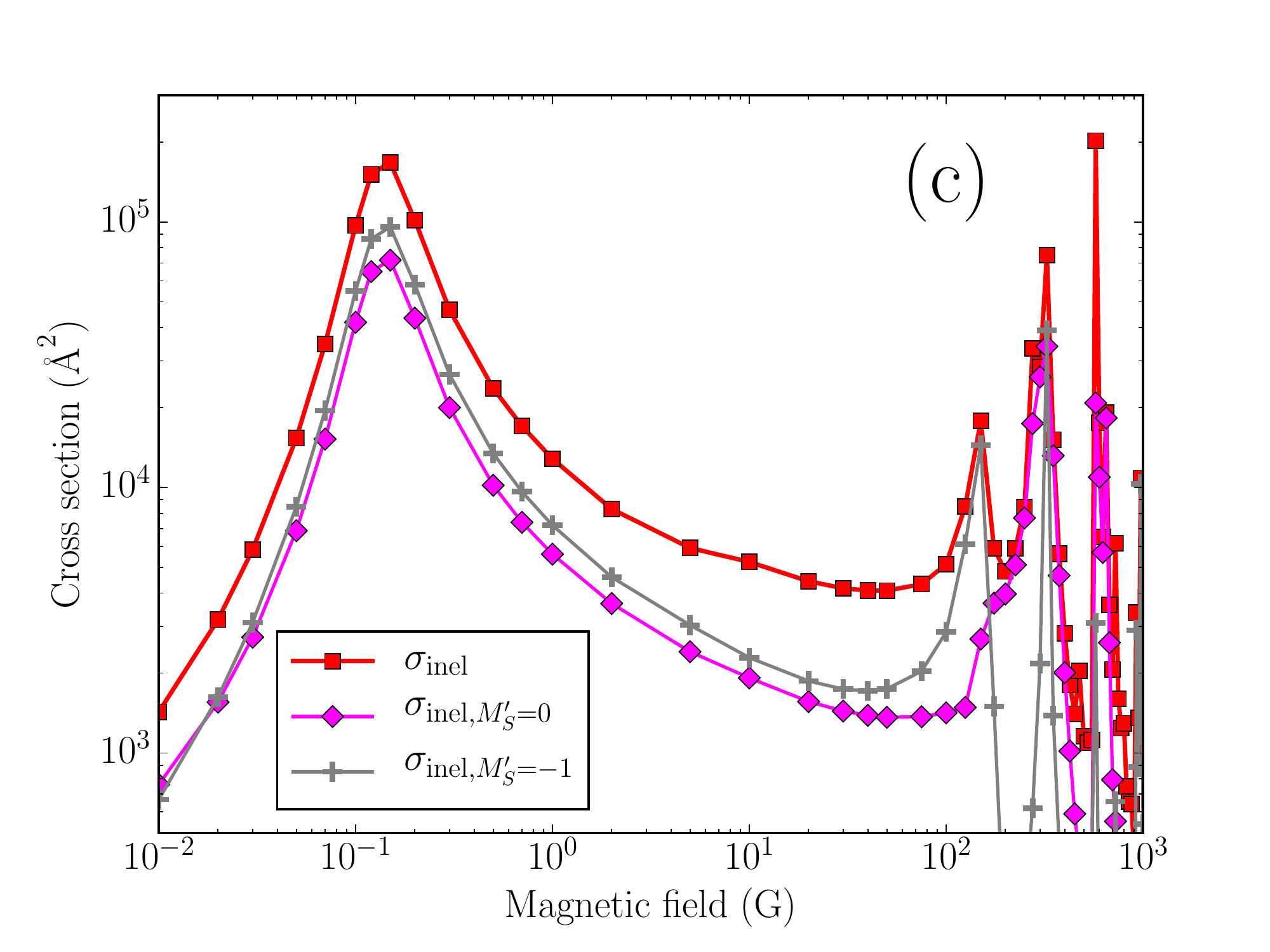}
\end{center}
\caption{(a) Magnetic field dependence of the elastic (circles) and inelastic (squares) cross sections calculated for the collision energy of $10^{-6}$ cm$^{-1}$. The inelastic cross sections calculated with the magnetic dipole-dipole interaction omitted are shown as stars. 
(b) Incoming partial wave decomposition of the inelastic cross section. (b) Final-state decomposition of the inelastic cross section. } 
\label{figure6}
\end{figure}

In Fig.~\ref{figure6}(a), we plot the magnetic field dependence of the cross sections for elastic scattering and spin relaxation in spin-polarized Rb-SrF collisions at a collision energy of  $10^{-6}$~cm$^{-1}$. We observe a broad resonance profile in the inelastic cross section centred at $B=0.1$~G, where inelastic scattering occurs 20 times faster than elastic scattering.   With further increase in magnetic field, the inelastic cross section decreases by more than an order of magnitude, whereas the elastic cross section remains essentially independent of the field. A dense and complicated resonance pattern emerges above $B=100$~G, where the ratio of elastic to inelastic   cross sections $\gamma$ varies rapidly from unity to above 100. Thus, it may be possible to enhance the efficiency of sympathetic cooling by tuning the inelastic cross sections with an applied magnetic field. 

Spin relaxation in ultracold collisions of $^2\Sigma$ molecules in their ground rotational states with $^2$S atoms is mediated by two mechanisms, direct and indirect \cite{pra11,RomanPRA03}. The direct mechanism is due to the long-range magnetic dipole-dipole interaction between the electronic spins of the collision partners given by the term $\hat{V}_{\mathrm{dd}}$ in Eq. (\ref{eq:Hdip}) \cite{pra11,pra17,JanssenPRA11,JanssenEPJD11}. The indirect mechanism is a combined effect of the intramolecular spin-rotation interaction and the coupling between the rotational states of the molecule induced by the anisotropy of the interaction potential \cite{pra11,RomanPRA03}.
Previous theoretical studies have found that spin-relaxation in Li~+~CaH and Li~+~SrOH collisions occur predominantly via a direct mechanism and that the indirect mechanism is strongly suppressed at low collision energies ($E_C< 10^{-3}$~cm$^{-1}$) \cite{pra11,pra17}. In order to compare these mechanisms for Rb~+~SrF collisions, we plot in Fig.~\ref{figure6}(a) the inelastic cross section calculated with the magnetic dipole-dipole interaction term $\hat{V}_{\mathrm{dd}}$ omitted from the scattering Hamiltonian. We observe a dramatic reduction of the spin relaxation cross section over the entire magnetic field range, except for a narrow resonance at $B=250$ G.


In order to further inspect the spin relaxation mechanisms, we show in Fig.~\ref{figure6}(b) the incoming partial wave contributions to the inelastic cross section. Below $B=300$ G,  the inelastic cross section is dominated by the incoming $p$-wave contribution. The incoming $s$-wave contribution becomes comparable in magnitude in the vicinity of scattering resonances.  The results plotted in Figs.~\ref{figure6}(a) and \ref{figure6}(b) allow us to conclude that spin relaxation in spin-polarized Rb~+~SrF collisions is driven by the magnetic dipole-dipole interaction between the electron spins of Rb and SrF.


As follows from Eq. (\ref{eq:Hdip}), the magnetic dipole-dipole interaction has non-zero matrix elements  between all of the $|M_{S_A}\rangle|M_{S_B}\rangle $ spin basis states. 
As a result, the magnetic dipole-dipole interaction can cause either single spin-flip relaxation, in which the electron spins of either Rb or SrF are flipped or double spin-flip relaxation, in which both of the electron spins are flipped.
The projection of the total electron spin of the Rb-SrF complex on the magnetic field axis $M_S=M_{S_A}+M_{S_B}$ changes by 1 in a single spin-flip transition ($|M_S=1\rangle \to |M_S'=0\rangle$) and by 2 in a double spin-flip transition ($|M_S=1\rangle \to |M_S'=-1\rangle$).
In contrast, the indirect mechanism mediated by the spin-rotation interaction \cite{RomanPRA03} can only change  the projection of the molecule's electron spin $M_{S_B}$, and thus only the single-flip $M_S=1 \to M_S'=0$ transition is allowed.

 Figure~\ref{figure6}(c) shows the final state-resolved inelastic cross sections for Rb-SrF collisions. We observe that double spin-flip relaxation  is slightly more efficient than single spin-flip relaxation at low magnetic fields. Interestingly, the double spin-flip relaxation occurs without changing the initial partial wave component, via the process $|M_{S}=1 \rangle |l=1,M_l=-1\rangle \to |M_{S}'=-1 \rangle |l'=1,M_l'=1\rangle$ within the ground rotational state manifold $(N=N'=0)$.





\section{Summary and conclusions}

We have presented an {\it ab initio} study of ultracold collisions in a heavy, spin-polarized mixture of Rb($^2$S) atoms and SrF($X^2\Sigma^+$) molecules in the presence of an external magnetic field. We developed an accurate {\it ab initio} interaction PES for the triplet $^3A'$ electronic state of Rb-SrF  using the state-of-the-art CCSD(T) method and large correlation-consistent basis sets.
The PES features a deep minimum and an extremely steep dependence on the Rb-SrF bending angle $\theta$, making the Rb-SrF interaction  strongly anisotropic.    Using the {\it ab initio} PES,  we carried out converged quantum scattering calculations using the total angular momentum representation in the BF coordinate frame \cite{jcp10}, demonstrating the feasibility of such calculations on heavy, strongly anisotropic atom-molecule collision systems.

 The inelastic collisions  change the value of the molecule's electron spin projection $M_S$ on the magnetic field axis, leading to magnetic trap loss. The ratio  $\gamma$ of elastic to inelastic collision rates is a key predictor of successful atom-molecule sympathetic cooling in a magnetic trap. Our calculations predict that ultracold spin-polarized Rb-SrF mixtures are relatively stable against collisional relaxation ($\gamma>10$) over most of the collision energy and magnetic field ranges explored in this work ($E_C=10^{-6}-10^{-3}$ cm$^{-1}$ and $B=0-1000$~G).  It is important to point out, however, that small changes in the Rb-SrF PES  can lead to dramatic variations of the scattering cross sections. Because the estimated uncertainty in our PES is about $5\%$, our scattering calculations presented in this paper should be considered as qualitatively accurate.   A detailed analysis of the effect of the uncertainties of the interaction potentials will be presented in future work \cite{CRDmodel}.  
 
 Our calculations predict a significant magnetic field dependence of the inelastic cross section at ultralow collision energies (see Fig.~\ref{figure6}), which suggests the possibility of tuning inelastic collision rates by applying an external  magnetic field to optimize the efficiency of sympathetic cooling, as suggested before for He-O$_2$ and Li-SrOH \cite{He-O2,pra17}.
The inelastic spin relaxation in cold Rb~+~SrF  collisions is mainly driven by a direct mechanism mediated by the magnetic dipole-dipole interaction between the electronic spins of Rb and SrF.

It is instructive to compare the collisional properties of Rb-SrF with those of the lighter collision systems Li-SrOH and Li-CaH explored in our previous work \cite{pra11,pra17}. While the potential depths and anisotropies are comparable in all of the alkali-molecule systems, the lighter reduced masses of Li-SrOH and Li-CaH result in higher centrifugal barriers. As~a~result,  the $s$-wave regime of Li-SrOH and Li-CaH collisions occurs at higher collision energies than that of Rb-SrF collisions. In addition, as mentioned in Sec. IIIA, the presence of a near-threshold $p$-wave resonance at low magnetic fields modifies the Wigner scaling of  Rb-SrF spin relaxation cross sections, making them almost independent of the collision energy [see Fig.~\ref{figure6}(b)]. In contrast, the spin relaxation cross sections for Li-CaH and Li-SrOH exhibit the expected $s$-wave Wigner scaling as $E_C\to 0$ with $\sigma_\text{inel}\propto E_C^{-1/2}$.   Finally, the resonance peaks in the magnetic field dependence of the spin relaxation cross sections for Rb-SrF are much narrower than those calculated previously for Li-SrOH \cite{pra17}. This suggests that the resonances in Rb-SrF collisions decay mainly by tunnelling through a $p$-wave centrifugal barrier in the incoming collision channel, whereas those in Li-SrOH collisions decay by a mechanism not involving tunnelling in the incoming channel.  

In future work, it would be interesting to explore the collisional properties of non-fully spin-polarized  initial states of Rb and SrF (which would require  explicit consideration  of the strongly attractive singlet PES) and elucidate the effects of hyperfine interactions on scattering observables  at low magnetic fields. 
Measurements of inelastic collision rates in ultracold Rb-SrF mixtures as a function of magnetic field would be desirable to constrain the interaction PES.


\section*{Acknowledgements}
We are grateful to David DeMille for stimulating discussions and to Roman Krems for generously sharing  his group's high-performance computing resources. This work was supported by NSF grant No. PHY-1607610. MK and PZ were supported by Grant No. DEC-2012/ 07/B/ST4/01347.  We are also grateful for the computer time provided by the Wroclaw Centre of Networking and Supercomputing (Project No. 218). 
 
\appendix*
\section{Basis set convergence}

In this section, we examine the convergence properties of the Rb~+~SrF cross sections with respect to the basis set truncation parameters $J_\mathrm{max}$ and $N_\mathrm{max}$, which determine the  maximum quantum numbers of the total angular momentum $J$ of the Rb-SrF collision complex and the rotational angular momentum $N$ of SrF.

\begin{figure}[ht]
\begin{center}
\includegraphics[height=0.35\textheight,keepaspectratio]{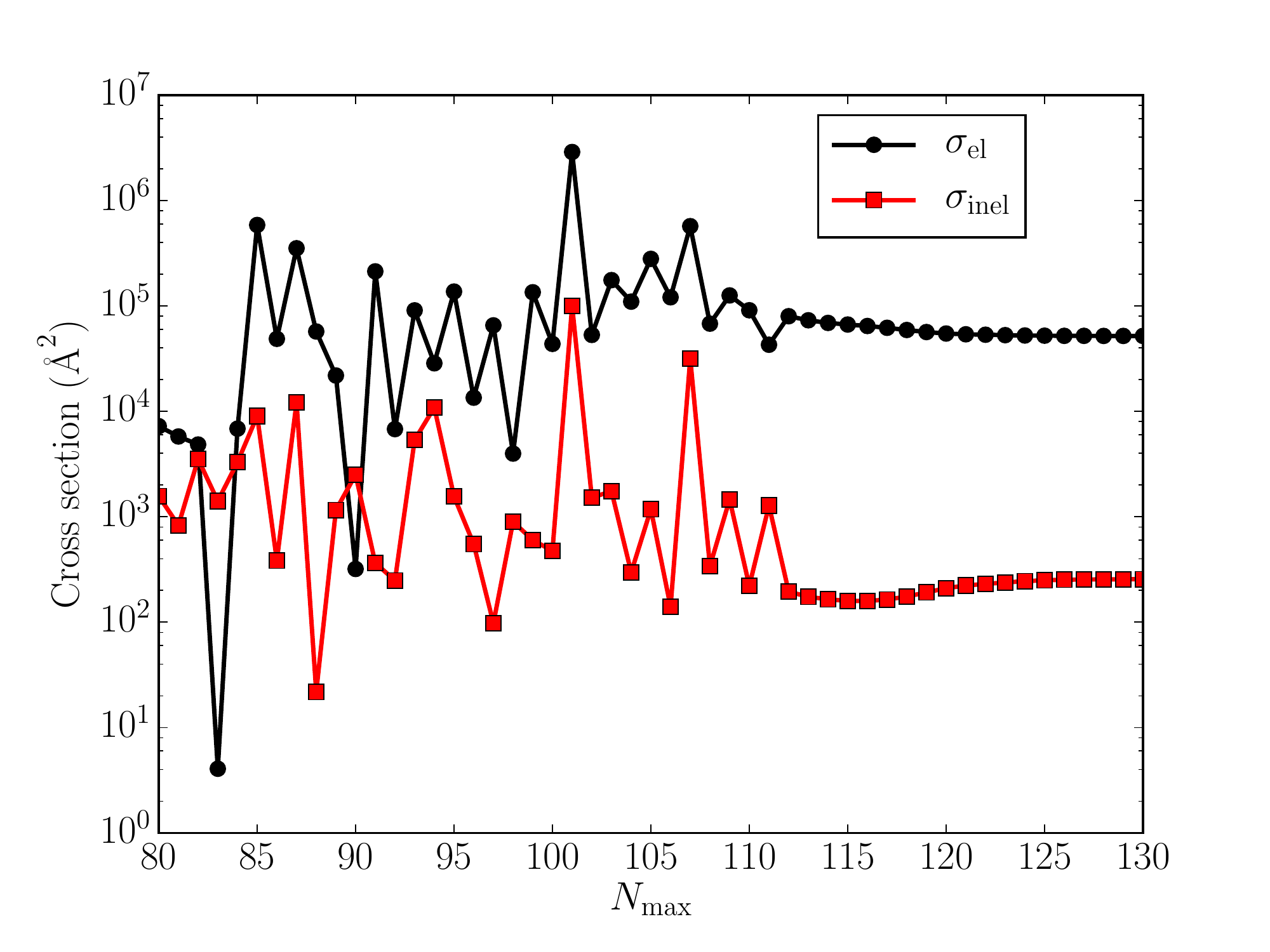}
\end{center}
\caption{Convergence of the elastic and inelastic cross sections with respect to the number of rotational states included in the basis set at the collision energy of $10^{-6}$ cm$^{-1}$. The magnetic field is 100 G and $J_\text{max}=1$. }
\label{figure2}
\end{figure}

The convergence of the elastic ($\sigma_\mathrm{el}$) and inelastic ($\sigma_\mathrm{inel}$) cross sections for fully spin-polarized Rb-SrF collisions with respect to the value of $N_\text{max}$ is shown in Fig.~\ref{figure2} for $B=100$~G, $J_\mathrm{max}=1$ and $E_C=10^{-6}$~cm$^{-1}$. The cross sections display rapid oscillations, which persist until $N_\text{max}\geq 110$, and we find that $N_\mathrm{max}=125$ is necessary to produce the cross sections converged to within 2.5\%.

\begin{figure}[ht]
\begin{center}
\includegraphics[height=0.35\textheight,keepaspectratio]{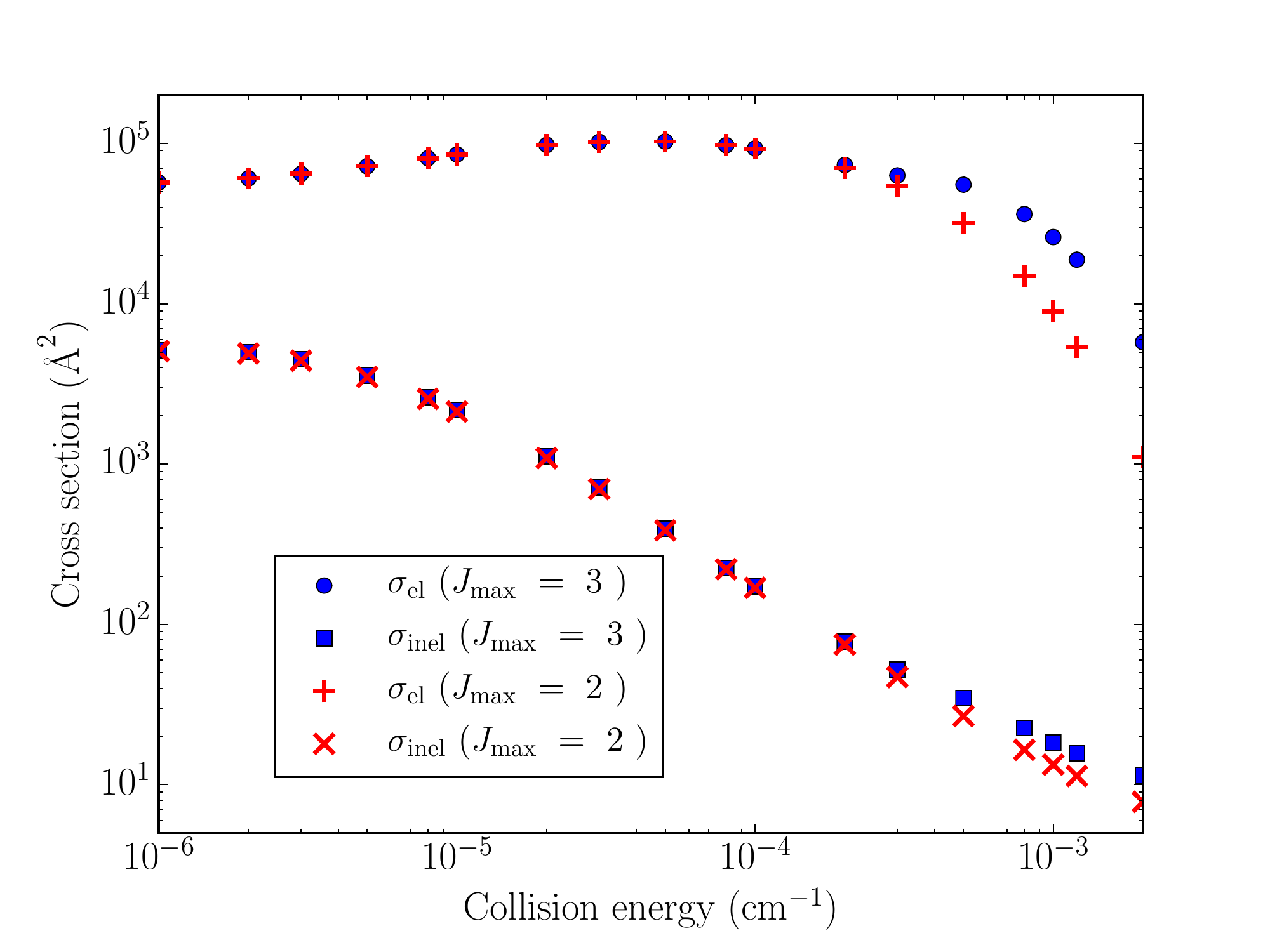}
\end{center}
\caption{Convergence of the elastic and inelastic cross sections with respect to the maximum total angular momentum value in the basis set: $J_\text{max}=3$ (circles and squares) and $J_\text{max}=2$ (pluses and crosses). The magnetic field is 100 G.}
\label{figure3}
\end{figure}

To examine the convergence with respect to the maximum value of the total angular momentum $J_\text{max}$, we plot $\sigma_\mathrm{el}$ and $\sigma_\mathrm{inel}$ as a function of collision energy in Fig. \ref{figure3} for $J_\text{max}=2$ and 3 at $B=100$ G. Adequate convergence is achieved with $J_\text{max}=2$ through the entire collision energy region.
As discussed previously \cite{VolpiBohn,Roman04,TutorialChapter2017,pra17}, indirect spin-relaxation in the incoming $s$-wave channel must be accompanied by a change of the orbital angular momentum from $l=0$  to $l=2$.
As a result, in order to properly describe the $d$-wave states in the outgoing collision channels, it is necessary to include at least 4 total angular momentum states  ($J_\text{max}\ge 3$) in the basis set. 
On the other hand, the incoming $p$-wave can make a transition to the outgoing $p$-wave by changing $m_l$, the projection of $l$ on the magnetic field axis. Thus, the $s$ and $p$-waves in the entrance and exit collision channels can be described by a smaller basis set with $J_\text{max}=2$. 
To properly account for all of the partial waves in the entrance and exit collision channels, we  choose to use $J_\mathrm{max}=3$ and $N_\text{max}=125$  for the production calculations.


\clearpage
\newpage

\clearpage
\newpage

\begin{thebibliography}{99}

\bibitem{njp09}
L. D. Carr, D. DeMille, R. V. Krems, and J. Ye, Cold and ultracold molecules: science, technology and applications, New J. Phys. {\bf 11}, 055049 (2009).

\bibitem{MishaMolPhys13}
M. Lemeshko, R. V. Krems, J. M. Doyle, and S. Kais, Manipulation of molecules with
electromagnetic fields, Mol. Phys. {\bf 111}, 1648 (2013).

\bibitem{JunARPC14}
B. K. Stuhl, M. T. Hummon, and J. Ye, Cold state-selected molecular collisions and reactions, Annu. Rev. Phys. Chem. {\bf 65}, 501 (2014).

\bibitem{RomanPCCP08}
R. V. Krems, Cold Controlled Chemistry, Phys. Chem. Chem. Phys. {\bf 10}, 4079-4092 (2008).


\bibitem{FaradayDiscuss09}
T. V. Tscherbul, G. Groenenboom, R. V. Krems, and A. Dalgarno, Dynamics of OH($^2\Pi$)-He collisions in combined electric and magnetic fields?, Faraday Discuss. {\bf 142}, 127 (2009).

\bibitem{prl06}
T. V. Tscherbul and R. V. Krems, Controlling electronic spin relaxation of cold molecules with electric fields, Phys. Rev. Lett. {\bf 97}, 083201 (2006).

\bibitem{KRb10}
S. Ospelkaus, K.-K. Ni, D. Wang, M. H. G. de Miranda,
B. Neyenhuis, G. Qu\'em\'ener, P. S. Julienne, J. L. Bohn, D. S. Jin, and J. Ye, Quantum-state controlled chemical reaction of ultracold KRb molecules, Science {\bf 327}, 853 (2010).

\bibitem{KRb10a}
K.-K. Ni, S. Ospelkaus, D. Wang, G. Qu\'em\'ener, B. Neyenhuis, M. H. G. de Miranda, J. L. Bohn, J. Ye, and D. S. Jin, Dipolar collisions of polar molecules in the quantum regime, Nature {\bf 464}, 1324 (2010).

\bibitem{KRbNaturePhys}
M. H. G. de Miranda, A. Chotia, B. Neyenhuis, D.Wang, G. Qu\'em\'ener, S. Ospelkaus, J. L. Bohn, J. Ye, and D. S. Jin, Controlling the quantum stereodynamics of ultracold bimolecular reactions, Nat. Phys. {\bf 7}, 502 (2011).

\bibitem{BrianBalaGeomPhase1}
B. K. Kendrick, J. Hazra, and N. Balakrishnan, The geometric phase controls ultracold chemistry, Nat. Commun. {\bf 6}, 7918 (2015).


\bibitem{QuantumChaos1}
J. F. E. Croft and J. L. Bohn, Long-lived complexes and chaos in ultracold molecular collisions, \pra{89}{012714}{2014}.

\bibitem{QuantumChaos2}
J. F. E. Croft, C. Makrides, M. Li, A. Petrov, B. K. Kendrick, N. Balakrishnan, S. Kotochigova, 
Universality and chaoticity in ultracold K+KRb chemical reactions, Nat. Commun. {\bf  8}, 15897 (2017).

\bibitem{prl15}
T. V. Tscherbul and R. V. Krems, Tuning Bimolecular Chemical Reactions by Electric Fields,  Phys.~Rev.~Lett. {\bf 115}, 023201 (2015).

\bibitem{Jonathan98}
J. D. Weinstein, R. deCarvalho, T. Guillet, B. Friedrich, and J. M. Doyle, Magnetic Trapping of Calcium Monohydride Molecules at Millikelvin Temperatures, Nature (London) {\bf 395}, 148 (1998).

\bibitem{NHtrapping}
W. C. Campbell, E. Tsikata, H.-I Lu, L. D. van Buuren, and J. M. Doyle, Magnetic Trapping and Zeeman Relaxation of NH($^3\Sigma^-$),  Phys. Rev. Lett. {\bf 98}, 213001 (2007).

\bibitem{OHtrappingMeijerARPC}
S. Y. T. van de Meerakker, P. H. M. Smeets, N. Vanhaecke, R. T. Jongma, and G. Meijer. Deceleration and electrostatic trapping of OH radicals. Phys. Rev. Lett. {\bf 94}, 023004 (2005).


\bibitem{O2trappinEd1}
N. Akerman, M. Karpov, Y. Segev, N. Bibelnik, J. Narevicius, and E. Narevicius, Trapping of Molecular Oxygen together with Lithium Atoms, \prl{119}{073204}{2017}.

\bibitem{CH3trapping}
Y. Liu, M. Vashishta, P. Djuricanin, S. Zhou, W. Zhong, T. Mittertreiner, D. Carty, and T. Momose, Magnetic Trapping of Cold Methyl Radicals, Phys. Rev. Lett. {\bf 118}, 093201 (2017).

\bibitem{Ivan16}
I. Kozyryev, L. Baum, K. Matsuda, B. L. Augenbraun, L. Anderegg, A. P. Sedlack, and J. M. Doyle, Sisyphus Laser Cooling of a PolyatomicMolecule, Phys. Rev. Lett. {\bf 118}, 173201 (2017).

\bibitem{SrFnature14}
J. F. Barry, D. J. McCarron, E. B. Norrgard, M. H. Steinecker, and D. DeMille, Magneto-optical trapping of a diatomic molecule, Nature (London) {\bf 512}, 286 (2014).

\bibitem{SrFthesisJohnBarry}
J. Barry, Laser cooling and slowing of a diatomic molecule, PhD thesis, Yale University, 2013.

\bibitem{SrFprl16}
E. B. Norrgard, D. J. McCarron, M. H. Steinecker, M. R. Tarbutt, and D. DeMille, Submillikelvin Dipolar Molecules in a Radio-Frequency Magneto-Optical Trap, \prl{116}{063004}{2016}.

\bibitem{JunYO}
M. T. Hummon, M. Yeo, B. K. Stuhl, A. L. Collopy, Y. Xia, and J. Ye, 2D Magneto-Optical Trapping of Diatomic Molecules, \prl{110} {143001}{2013}.

\bibitem{CaFbelowDopplerLimit} 
S. Truppe, H. J. Williams, M. Hambach, L. Caldwell, N. J. Fitch, E. A. Hinds, B. E. Sauer and M. R. Tarbutt, Molecules cooled below the Doppler limit, Nature Phys. {\bf 13}, 1173 (2017).

\bibitem{ChinReview}
C. Chin, R. Grimm, P. Julienne, and E. Tiesinga, Feshbach resonances in ultracold gases, Rev. Mod. Phys. {\bf 82}, 1225 (2010).

\bibitem{LaraRb-OHprl}
M. Lara, J. L. Bohn, D. Potter, P. Soldan, and J. M. Hutson, Ultracold Rb-OH collisions and prospects for sympathetic cooling, {Phys. Rev. Lett.} {\bf 97}, 183201 (2006).


\bibitem{Science01}
G. Modugno, G. Ferrari, G. Roati, R. J. Brecha, A. Simoni, and M. Inguscio, Bose-Einstein condensation of potassium atoms by sympathetic cooling, Science {\bf 294}, 1320 (2001).

\bibitem{JosephDFG06}
S. Aubin, S. Myrskog, M. H. T. Extavour, L. J. Leblanc, D. Mckay, A. Stummer, and J. H. Thywissen, Rapid sympathetic cooling to Fermi degeneracy on a chip, Nat. Phys. {\bf 2}, 384 (2006).

\bibitem{VolpiBohn}
A. Volpi and J. L. Bohn, Magnetic-field effects in ultracold molecular collisions, Phys. Rev. A {\bf 65}, 052712 (2002).

\bibitem{Roman04}
R. V. Krems and A. Dalgarno, Quantum-mechanical theory of atom-molecule and molecular collisions in a magnetic field: Spin depolarization,   J. Chem. Phys. {\bf 120}, 2296 (2004).

\bibitem{Mg-NH}
A. O. G. Wallis and J. M. Hutson, Production of ultracold NH molecules by sympathetic cooling with Mg, \prl{103}{183201}{2009}.

\bibitem{TutorialChapter2017}
T. V. Tscherbul, Effects of External Magnetic Fields on Cold Molecular Collisions, in {\it Low Energy and Low
Temperature Molecular Scattering}, ed. by A. Osterwalder and O. Dulieu (Royal Society of Chemistry), in press (2017).


\bibitem{HutsonTarbutt}
J. Lim, M. D. Frye, J. M. Hutson, and M. R. Tarbutt, Modeling sympathetic cooling of molecules by ultracold atoms, \pra{92}{053419}{2015}.


\bibitem{NHisotopes}
W. C. Campbell, T. V. Tscherbul, H.-I Lu, E. Tsikata, R. V. Krems, and J. M. Doyle, Mechanism of collisional spin relaxation in $^3\Sigma$ molecules, Phys. Rev. Lett. {\bf 102}, 013003 (2009).

\bibitem{N-NH}
M.~T. Hummon, T.~V. Tscherbul, J. K{\l}os, H.-I Lu, E. Tsikata,  W.~C. Campbell, A. Dalgarno, and J.~M. Doyle, Cold N~+~NH Collisions in a Magnetic Trap,  {Phys. Rev. Lett.} {\bf 106}, 053201 (2011).

\bibitem{PiotrN-NH}
P. {\.Z}uchowski and J. M. Hutson, Cold collisions of N atoms and NH molecules in magnetic fields, Phys. Chem. Chem. Phys. {\bf 13}, 3669-3680 (2011).

\bibitem{HutsonH-OH}
M. L. Gonz{\'a}lez-Mart{\'i}nez and J. M. Hutson, Ultracold Hydrogen Atoms: A Versatile Coolant to Produce Ultracold Molecules, \prl{111}{203004}{2013}.

\bibitem{RomanPRA03}
R. V. Krems, A. Dalgarno, N. Balakrishnan and G. C. Groenenboom, Spin-flipping transitions in ${}^{2}\ensuremath{\Sigma}$ molecules induced by collisions with structureless atoms, Phys. Rev. A {\bf 67}, 060703 (2003).


\bibitem{pra11}
T. V. Tscherbul, J. K{\l}os, and A. A. Buchachenko, Ultracold spin-polarized mixtures of $^2\Sigma$ molecules with $S$-state atoms: Collisional stability and implications for sympathetic cooling, Phys. Rev. A {\bf 84}, 040701(R) (2011).

\bibitem{pra17}
M. Morita, J. K{\l}os, A.A. Buchachenko, and T. V. Tscherbul, Cold collisions of heavy  $^2 \Sigma$  molecules with alkali-metal atoms   in a magnetic field: {\it Ab initio} analysis and prospects for sympathetic    cooling of SrOH($^2\Sigma$) by Li($^2$S),  Phys. Rev. A {\bf 95}, 063421 (2017).


\bibitem{JonathanLiCaH}
V. Singh, K. S. Hardman, N. Tariq, M.-J. Lu, A. Ellis, M. J. Morrison, and J. D. Weinstein, ``Chemical Reactions of Atomic Lithium and Molecular Calcium Monohydride at 1 K'', \prl{108}{203201}{2012}.

\bibitem{PiotrJPCA17}
M. B. Kosicki, D. K\c{e}dziera, and P. S. {\.Z}uchowski, Ab Initio Study of Chemical Reactions of Cold SrF and CaF Molecules with Alkali-Metal and Alkaline-Earth-Metal Atoms: The Implications
for Sympathetic Cooling, J. Phys. Chem. A {\bf 121}, 4152 (2017).

\bibitem{N2}
T. V. Tscherbul, J. K{\l}os, A. Dalgarno, B. Zygelman, Z. Pavlovic, M. T. Hummon, H.-I. Lu, 
E. Tsikata, and J. M. Doyle, Collisional properties of cold spin-polarized nitrogen gas: Theory, experiment, and prospects as a sympathetic coolant for trapped atoms and molecules, \pra{82}{042718}{2010}

\bibitem{jcp07}
E. Abrahamsson, T. V. Tscherbul, and R. V. Krems, Inelastic collisions of cold polar molecules in nonparallel electric and magnetic fields, {J. Chem. Phys.} {\bf 127}, 044302 (2007).



\bibitem{JacekPRA15} M. Warehime and J. K{\l}os, Nonadiabatic collisions of CaH with Li: Importance of spin-orbit-induced spin relaxation in spin-polarized sympathetic cooling of CaH, Phys. Rev. A {\bf 92},  032703 (2015).

\bibitem{JanssenN-NH}
L. M. C. Janssen, A. van der Avoird, and G. C. Groenenboom, Quantum Reactive Scattering of Ultracold NH($X^3\Sigma^-$) Radicals in a Magnetic Trap, \prl{110}{063201}{2013}.


\bibitem{jpb16}
Yu. V. Suleimanov and T. V. Tscherbul, Cold NH-NH collisions in a magnetic field: Basis set convergence versus sensitivity to the interaction potential, J. Phys. B {\bf 49}, 204002  (2016).


\bibitem{Hampel:92}
C. Hampel, K. Peterson, H.-J. Werner,  A comparison of the efficiency and accuracy of the quadratic configuration interaction (QCISD), coupled cluster (CCSD), and Brueckner coupled cluster (BCCD) methods, Chem. Phys. Lett. {\bf 190}, 1 (1992).

\bibitem{Knowles:94}
M. J. O. Deegan, P. J. Knowles, Perturbative corrections to account for triple excitations in closed and open shell coupled cluster theories, Chem. Phys. Lett. {\bf 227}, 321 (1994).

\bibitem{MOLPRO_brief:2015}
H.-J. Werner, P. J. Knowles, G. Knizia, F. R. Manby, M. {Sch\"{u}tz}, P. Celani, W.{Gy\"orffy}, D. Kats, T. Korona, R. Lindh, \textit{et al.}, {\textit{MOLPRO}, version 2015.1: A package of ab initio programs}, {see http://www.molpro.net} (2015). 

\bibitem{Lim:2005}
I. S. Lim, P. Schwerdtfeger, B. Metz, H. Stoll, All-electron and relativistic pseudopotential studies for the group 1 element polarizabilities from K to element 119, J. Chem. Phys. {\bf 122}, 104103 (2005).

\bibitem{Sr_ECP28MDF} 
I.S. Lim, H. Stoll, P. Schwerdtfeger, Relativistic small-core energy-consistent pseudopotentials for the alkaline-earth elements from Ca to Ra, J. Chem. Phys. {\bf 124}, 034107 (2006).

\bibitem{Boys:70} 
S. F. Boys and F. Bernardi, The calculation of small molecular interactions by the differences of separate total energies. Some procedures with reduced errors, Mol. Phys. {\bf 19}, 553 (1970)

\bibitem{Janssen:2009}
L. M.~C. Janssen, G.~C Groenenboom, A.~Van Der~Avoird, P.~S.  Zuchowski, and R. Podeszwa, Ab initio potential energy surfaces for NH($^3\Sigma^-$)-NH($^3\Sigma^-$) with analytical long range
J. Chem. Phys. {\bf 131}, 224314 (2009).

\bibitem{Lee:1989}
T. J. Lee and P. R. Taylor, A diagnostic for determining the quality of single-reference electron correlation methods, Int. J. Quantum Chem., Quant. Chem. Symp., {\bf S23}, 199 (1989).

\bibitem{Huber}
K. P. Huber, G. Herzberg, {\it Molecular Spectra and Molecular Structure, IV. Constants of Diatomic Molecules} {published by Van Nostrand Reinhold}, (1979).

\bibitem{Ho:1996}
T. S. Ho and H. Rabitz, A general method for constructing multidimensional molecular potential energy surfaces from {\em ab initio} calculations, J. Chem. Phys. {\bf 104}, 2584 (1996).






























\bibitem{jcp10} 
T. V. Tscherbul and A. Dalgarno, Quantum theory of molecular collisions in a magnetic field: Efficient calculations based on the total angular momentum representation,  J. Chem. Phys. {\bf 133}, 184104 (2010).


\bibitem{jcp12a}
Yu.~V. Suleimanov, T. V. Tscherbul, and R. V. Krems, Efficient method for quantum calculations of molecule-molecule scattering properties in a magnetic field,  J. Chem. Phys. {\bf 137}, 024103 (2012). 











\bibitem{prop_1}
D. E. Manolopoulos, An improved log derivative method for inelastic scattering, J. Chem. Phys. {\bf 85}, 6425 (1986).

\bibitem{prop_2}
M.H. Alexander and D.E. Manolopoulos, A stable linear reference potential algorithm for solution of the quantum close-coupled equations in molecular scattering theory, J. Chem. Phys. {\bf 86}, 2044 (1987).


\bibitem{He-O2}
J. M. Hutson, M. Beyene and M. L. Gonz\'alez-Mart\'{\i}nez, Dramatic Reductions in Inelastic Cross Sections for Ultracold Collisions near Feshbach Resonances, Phys. Rev. Lett. {\bf 103}, 163201 (2009).


\bibitem{JanssenPRA11}
L. M. C. Janssen, P. S. {\.Z}uchowski, A. van der Avoird, G. C. Groenenboom, and J. M. Hutson, Cold and ultracold NH-NH collisions in magnetic fields, \pra{83}{022713}{2011}.

\bibitem{JanssenEPJD11}
L. M. C. Janssen, A. van der Avoird, and G.C. Groenenboom, On the role of the magnetic dipolar interaction in cold and ultracold collisions: Numerical and analytical results for NH($^3\Sigma$) + NH($^3\Sigma$), Eur. Phys. J. D {\bf 65}, 177 (2011).

\bibitem{CRDmodel}
M. Morita, R. V. Krems, and T. V. Tscherbul, to be published.



\end{thebibliography}
\end{document}